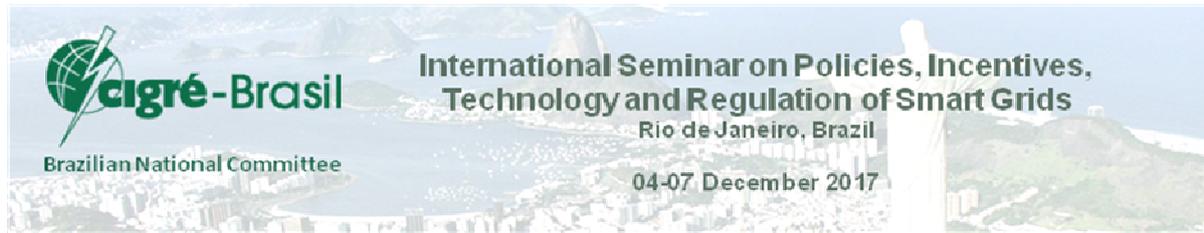

# Distribution Power Network Reconfiguration in the Smart Grid

Eonassis O. Santos, Joberto S. B. Martins

Salvador University – UNIFACS, Brazil

**SUMMARY**

The power network reconfiguration algorithm with an "R" modeling approach evaluates its behavior in computing new reconfiguration topologies for the power grid in the context of the Smart Grid. The power distribution network modelling with the "R" language is used to represent the network and support computation of different algorithm configurations for the evaluation of new reconfiguration topologies. This work presents a reconfiguration solution of distribution networks, with a construction of an algorithm that receiving the network configuration data and the nodal measurements and from these data build a radial network, after this and using a branch exchange algorithm And verifying the best configuration of the network through artificial intelligence, so that there are no unnecessary changes during the operation, and applied an algorithm that analyses the load levels, to suggest changes in the network.

**KEYWORDS**

Smart Grid, Distribution Network Reconfiguration, Distribution Network Reconfiguration Algorithm, Intelligent Network, "R" Language, Modeling, Artificial Intelligence, Machine Learning, HATSGA.

eonassis@msn.com

I – INTRODUCTION

The electrical networks are composed by stages of generation, transmission and distribution of electricity, these networks allow the loads to meet the generation points. These interconnections in their majority grid are made hierarchically and with changes in their configuration made manually and intuitively, that is without planning. In my work I will focus only on distribution networks.

Distribution networks are mostly radial type networks, since there are many facilities from an engineering point of view to operate a network in this way, it could be even more interesting if this network were mesh type, but the implementation costs would increase gigantically. The radial distribution networks are nothing more than a minimum fixed-base (substation) generating tree, which is reconfigured with keys, manually or remotely controlled, to achieve several possibilities of network reconfigurations, which may occur due to faults in some point of the network, preventive maintenance and to solve problems of overloading of feeders or substations.

Distribution networks with the expansion of Smarts-Grids technologies have deployed keys with remote controls that allow reconfiguration of the network remotely. These keys can also provide network data such as power flow, voltage and circuit impedance.

The distribution networks (RD) of electric power have configurations and loads of the most varied, each point of power of the network or substations of lowering has connected him commercial, residential, industrial and rural loads. Each type of load has its behavior as well as its traffic profile, the radial configuration imposed to this type of network aims to offer quality through the observance of limits of tension and observers of continuity of the service.

By altering topology and restoring supply or isolating faults, it is also possible to logistic operations in the network as transfer of loads between feeders, in turn balancing loads and relieving overloaded feeders, with this type of maneuver and possible to increase reliability levels of RD.

The problem is to find among the combinations of possible configurations of the network, a configuration that is the best for the whole system, where feeders, conductors, transformers and substations work in their best configuration while maintaining their radial configuration, reducing energy losses, consumers. It also protects the voltage levels of the load points, maximum current flow in conductors and transformers.

In this paper we will discuss an R-based network reconfiguration solution in the context of Smart Grid's networks, proposing a flexible model based on machine learning to be able to take increasingly flexible actions in a dynamic generation environment , making the RD protected and optimized, with a construction of an algorithm that receives the configuration data of the network and the nodal measurements and from this data build a radial network, after this and using a branch exchange algorithm and checking the best network configuration through linear regression.

The Simulation and testing environment was implemented in R language, and the Newton-Rapson method for calculating power flow, the network where the algorithm was tested are IEEE-14 BUS test networks.

II – MOTIVATION

Smart Grids (SG) networks are the solution to increasingly different demands being part of a network, intelligence in these types of networks and important so that we can respond with the required demand and operation. Some motivations are listed below:
- Common problem for RD's engineers;
- Benefit from a flexible and problem-solving computer approach;

• Find a solution to help in the decision process of real-time network reconfiguration.

In the context of Smart Grids Networks the modelling of the problem of reconfiguration of RDs based on "R" aims to help in the decision making of the problem of reconfiguration of RDs, the possible scenarios are:
• Optimization in the supply and use of RD resources, reconfiguring the network and doing load balancing.
  o Identifying points of attention and overload.
  o Proposing improvement in quality of service, respecting constraints of voltage levels, line power flow capacity and transformers capacity.
• Distributed Generation (GD) - Network reconfiguration with load balancing.
  o Identify sources of distributed energy through variations in load or demand in the DR.
  o Change the network so that it has the best possible configuration, including the GD points present in the network to optimize the load distribution.

III - MODELING THE PROBLEM

This work clearly fits into a graph learning problem, that there are several strands and ways to solve it, the problem will be approached considering the part between two network keys directly connected as if it were a vertex and the key as if it were the as shown in figure 1.

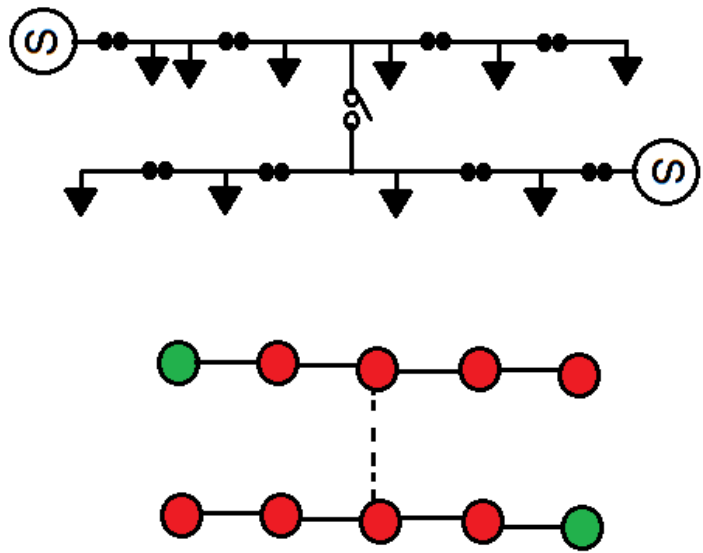

Figure 1 - Modeling of a distribution network for a graph.

In figure 1 we have an example of a typical minimum configuration of a distribution network with two feed points, the feeder and the root of our generating tree, the connected edges and the NC keys and the unconnected edges and the keys NA, any load of the circuit including the load points of the segment and represented by a load vertex, which comprises the load of the net present from one side of a key to the other side of the other key.
The initial problem is to create a sub graph from the graph of the main network so that the configuration of the interconnection of the branches is as efficient as possible, ie with less level of voltage drop, better distribution of loads between the feeders respecting the limit of each, notably a minimal tree problem with root restriction in a graph.

After creating the minimum generating tree, now is the time to improve its configuration through the minimization function applied to the sum of the loads and network impedance in the feeders. This point aims to improve the configuration of the network through branch exchange, where you get better configuration, keep the change, if it worsens the result, return to the previous state. Figure 2 demonstrates one of the benefits that we can have with the reconfiguration of RD as a function of time, scenario 1 at a given moment the feeders were overloaded, which is undesired condition of the system, already in scenario 2 through reconfiguration of the network keys a better distribution of the load between the feeders is obtained, there being no unwanted network overloads and feeders.

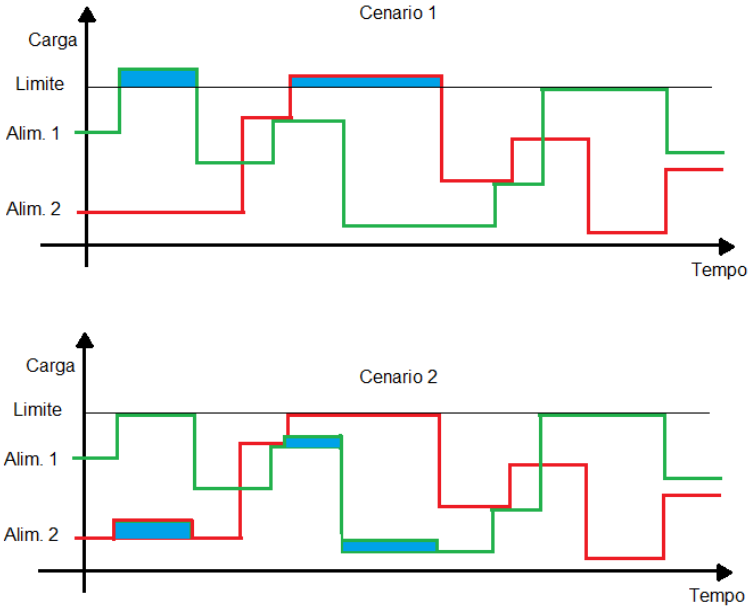

Figure 2 - Overflow in Feeders

IV - LINEAR REGRESSION MODEL

The artificial intelligence is already widely used in systems whose deterministic algorithms do not bring the solution, a RD and an example of a model where the solution is not static, it has variation in time and space, in front of this the proposal of this model and use learning of a machine to deal with the RD reconfiguration problem, the type of machine learning that best fits the problem and the regression, through a linear cost function that applies RD.
In order for the problem to be solved, the target value must always be the minimum among the analyzed feeders, in order to obtain the best possible network configuration.
The first step in the process of constructing a linear optimization model and the definition of the objective function and constraints of the problem. The criteria to be taken into account in this case will be load balancing, reduction of power losses, improvement of voltage levels and improvement in network reliability. The restrictions are linked to the regulation of Brazilian bodies, which define adequate levels.
With strategic focus of the electric power utilities, this work defines the following objective function (FO), based on the FO of the HATSGA algorithm (Calhau 2016) with necessary changes, for the problem of reconfiguration of distribution networks, under normal according to equation 1:

$$\text{FO:min } f = \sum_{i=0}^{n} k_i r_i \frac{P_i^2 + Q_i^2}{v_i^2} \cdot \Delta t \qquad (1)$$

Being:
i = Edge index;
n = Total number of edges;
r_i = Edge resistance i (Ω);
Q_i = Reactive power of the edge i;
P_i = Active power of the edge i;
v_i = Voltage of the main end of the edge i;
k_i = branch i, 1 and equal to closed, and 0 and equal to open; and
Δt = time interval (h) of the loading platform.

s.a.:
- Maintain the radial network;
- The feeders must not operate with overload;
- Voltage and current limits should not be violated; and
- During the maneuvers the voltage and current limits must not be violated.

V - MINIMUM GENERATING TREE

Among the algorithms of minimal tree we have some classics such as Kruskal and Prim, but for this problem we will use a variation of these, where the algorithm considers the network to be totally interconnected, that is, the fully connected graph, all the keys in the state NF, at this moment the algorithm must create several fixed base trees first defining the base that are the feeders, after that it is selecting edges with greater power flow and inserting in the trees one at a time, so that in the end I have the number of trees equal to the number of feeders of the network, thus making several radios, each connected to its feeder.

1. Choose feeder vertices to start the subgraphs
2. while there are vertices that are not in the subgraphs
3. Select an edge with higher power flow
4. Insert the edge with the highest power flow and its vertex in the subgraph

Algorithm 1 - Minimum Generating Tree Algorithm

VI - BRANCH EXCHANGE ALGORITHM (Branch Exchange)

This method consists in performing successive openings and key closures, considering that each in the network can be a possible solution to the problem, in RD's, the purpose of this method and to keep the RD always with the radial configuration.
In the algorithm proposed by (Pfitscher, 2013), the algorithm chooses one of the NA keys of the branch places it in NF position, from this moment the feeders are interconnected in parallel, at the moment a key is chosen closer to the maneuvered key in a the power flow, objective function of the new configuration, and it is checked if there was no violation of restrictions, if there is an improvement in the objective function of the two feeders, the steps above to the same side, if there is no improvement, return the previous step and move to the other direction, this process is repeated until you have the best possible network configuration, observed through the objective function.

VII – MODELING WITH "R"

The computational modeling was chosen to simulate the proposed algorithm, with this we were able to compare methods and simulate results before applying it to the real world, the "R" language was chosen because it is a free software development environment for computing, and optimized statistics, oriented and non-oriented data models, linear and nonlinear data models, multiple configurations, and flexibility in data handling and manipulation, as well as well-advanced power flow models and artificial intelligence models developed by language. The "R" works well with modeling graphs through libraries. The power flow computation simulations scripts (Gauss-Siedel and Newton-Raphson techniques) were implemented by RPowerLabs also using "R" and were added as a module of the algorithm.

To represent a RD, the most coherent way of assimilating would be with the graphs, a RD includes a set of elements that can be expressly represented by graphs, such as: NA keys and NF keys, distributed generation points and radial implementation mode. What can be widely implemented with features included in the "R" language.

The representation is part of a graph, this graph has vertices representing the points and load and measurement, the edges represent the connection branches, which may or may not be connected to the active network. In our case we have in Figure 17 an example of an IEEE-14BUS network mapping.

In distribution networks the connection lines have two states for bus interconnection, it can be closed or NF (normally closed), open or NA (normally open), this switch connects or not between the buses, it is present in all electrical systems.

In graph theory an RD is mapped generically as a connected graph G = {vertices, edges), where the buses are the vertices and the keys are the edges (Figure 3), through this representation we can include loops between vertices, and represent with flexibly the topologies.

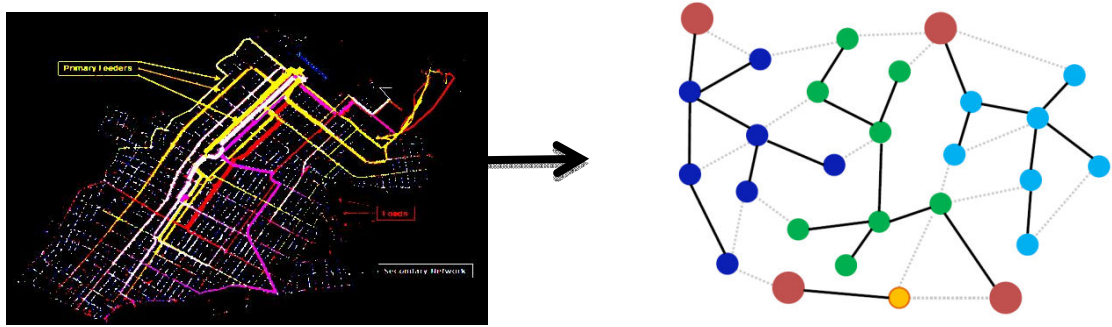

**Figure 3 - Mapping of RD in a Generic Graph**
**Source: D. Deka, S. Backhaus, MC, 2015**

The modeling of distribution networks in the language "R", and formed basically by vertices edges and the relation between them, that at the end of the process generates a generic graph representing the RD computationally, with the parameters represented through a matrix of data with information of each vertex, and another matrix with information of each edge of the system.

The connection keys and lines are represented by edges, for each edge of the graph we have an input with data in the matrix of parameters of edges, the lines of connection with any key are modeled with different types of edges, be it disconnector or junction. The degree of a vertex is given by its number of edge connections associated with it. The Matrix of Edge parameters is composed of n-dimensional vectors or a matrix represented in the "R", each entry in the matrix stores data parameters or attributes associated with the edge or the bus that it composes, including also its behavior when connected or not, whether it be downgrade, distributed generator or not and load bus. In the matrix we have all parameters relevant or not for computational simulation, besides data crucial for reconfiguration, we can also store operational data, including usage history or edge behavior. Because it is a data matrix it can be easily updated or adjusted or added information, any adjustment that adds or removes faces in the matrix changes its size and / or its structure accordingly.

For each vertex we have the type of bus that can be downslope, generator, load, generators and feeders, are modeled with different types in the matrix, in the matrix also we have voltage, resistance, reactance, capacitance, working angle, loads (P and Q), generation (P and Q), maximum and minimum bus limits, active and reactive power limits and parameters that define a short-cycle. The vertex parameter matrix stores transmission line data, which is used by the objective function (FO) to perform the linear regression calculations, parameters such as: Resistance, reactance, voltage and line limits. These RD electrical properties are crucial for FO calculation and network reconfiguration efficiently. Parameters such as Resistance and reactance are electrical properties of the line or bus, already the maximum and minimum limits of the bus are the thermal classification of the same and the derivation index represents the opening of the transformer. The matrix of vertices is equivalent to the matrix of edges and very relevant for the correct functioning of the machine learning algorithm.

Figure 4 shows the IEEE-14BUS bus system, this network and actual data provided by the IEEE for studies have 14 buses numbered from 1 to 14 and 20 links that are the connections between the buses, it also has 2 generators (G) and 3 synchronous compensators (C), which will be used in the simulation. Figure 5 shows the corresponding modeling of the IEEE-14BUS distribution network in a graph.

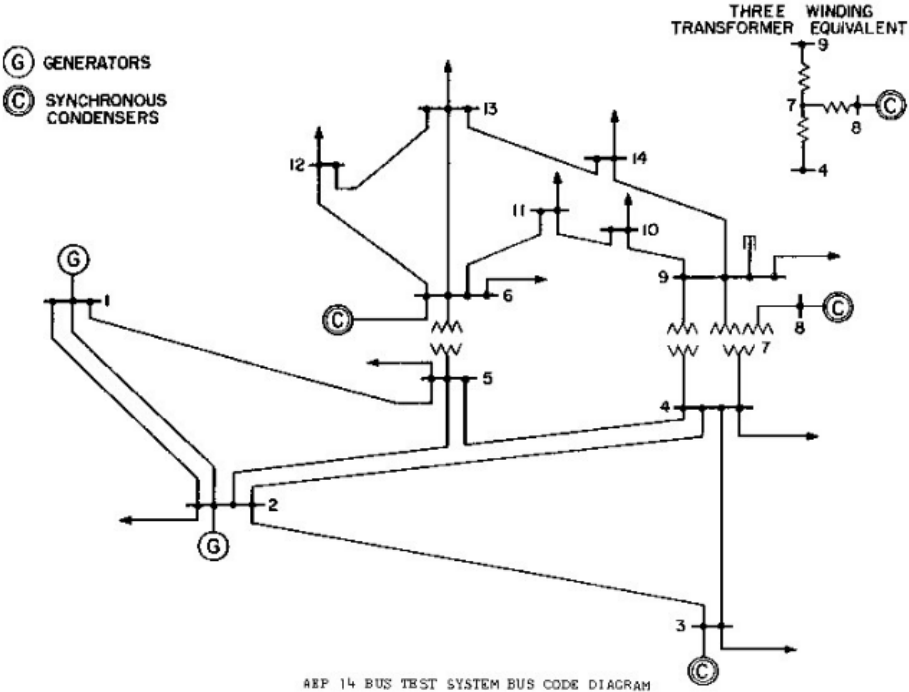

Figure 4 – IEEE-14 BUS Network
Source: IEEEXBUS

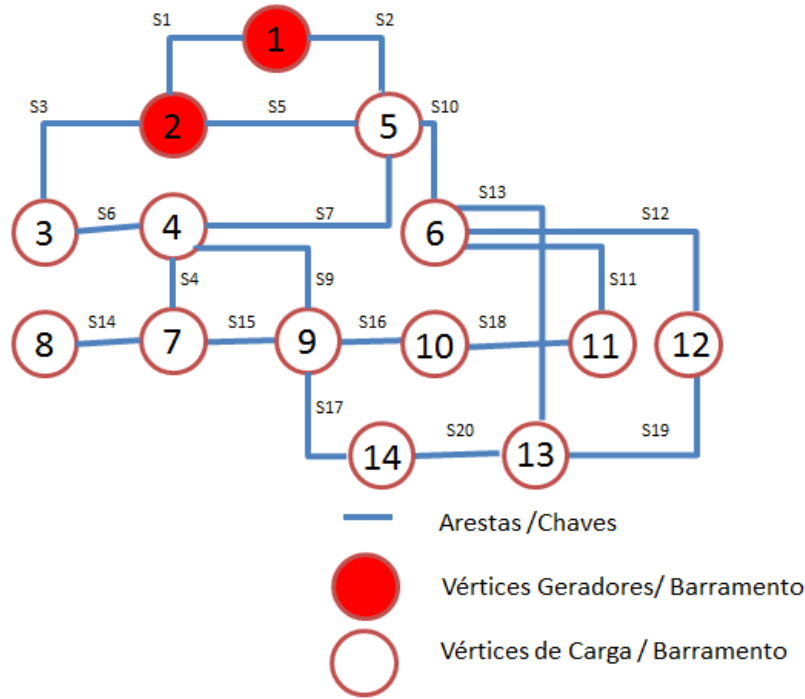

Figure 5 - Computational Model (graph) of the IEEE-14BUS

The objective function of the linear regression algorithm described in section V of the article has as objective a minimum loss of power, since it is an adaptation of the 2016 HATSGA algorithm, yet it allows configurations to use other objectives with the utilization of priority, useful functionality in cases where the concessionaire wants to set higher priorities for certain areas of its network.

The use of machine learning along with the branch exchange algorithm as mentioned in section VII reduces the space to search for a solution, eliminating configurations close to those not ideal for network, through linear regression the artificial intelligence algorithm always searches for the least number of the objective function as stated in section V, these two important steps already eliminate configurations that do not comply with network constraints, such as radial configuration, run-limit limits, resistance, reactance, among others.

The proposed algorithm seeks efficiency using linear regression and branch exchange techniques, taking the concept of spatial and temporal locus to the letter, used in all the large systems, where it establishes that if there was already a previous optimal solution with a certain configuration, the next best solution to the new demand of the network is close to the previous one, the branch exchange algorithm searches the spatial locality principle.

RD's usually have a radial topology. The proposed solution imposes constraints on the radiality of the tree generated with fixed base, the summary of the steps of the algorithm follows:

- The algorithm starts with a radial tree topology formed with fixed base, based on the network generation points, obtained through the modified PRIM algorithm to work with RD's, as explored in section VI.
- From this initial topology, the objective function value, which is the sum of the objective functions of each generating unit as explained in section V, is searched through the linear regression model, we now have the topology data and the data of face vertex and edge of the network loaded in their respective arrays.

- At this moment the algorithm of branch exchange is started, seeking efficiency in the losses and improvement in the objective function as explained in section VII.
- When the network restrictions, the reconfiguration will start from step 3.

VIII – EXPERIMENT

The algorithm starts with a representation of the topology modeled by a graph. In this example, we will use IEEE 14-BUS, supported by the data mapping of the data matrix of vertices and edges in "R" language as described in section IX (Figure 6). The basic process starts requirement and a radial topology, in our case multi-base trees, Figure 7 shows an example of an initial topology obtained from the IEEE14-BUS graph using the mapping via "R" using the first modified Prim´s algorithm explained in section VI, the Prim´s algorithm generates minimum trees based on the generators in the case vertices 1 and 2.

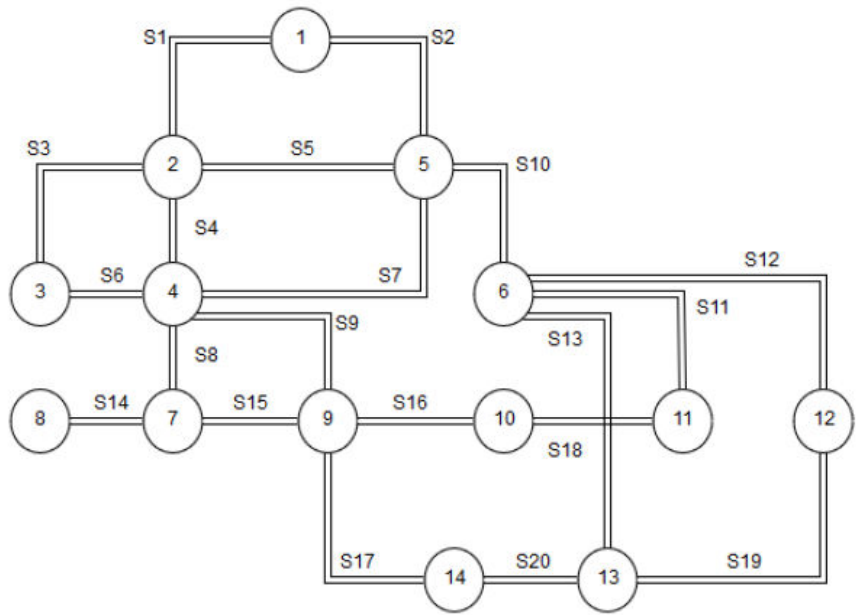

Figure 6 - Mapped through the "R" language of the IEEE-14BUS network.

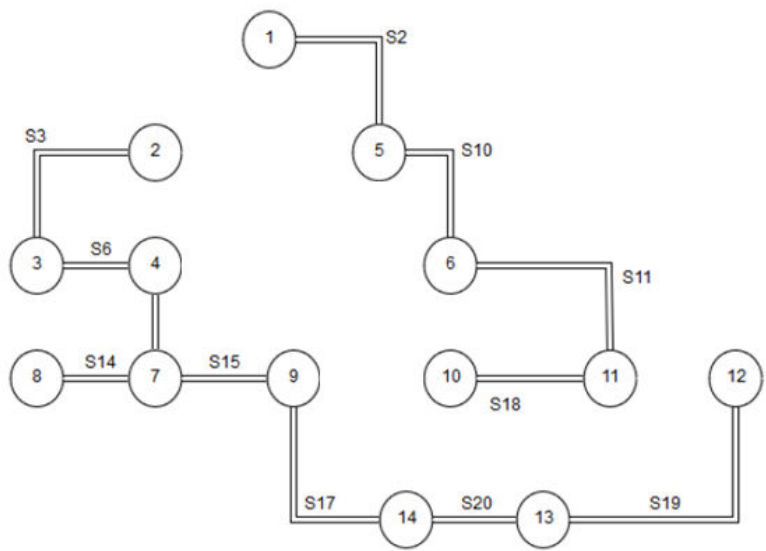

Figure 7 - Trees generated from the modified Prim´s algorithm

The generated sub-graphs have a list of open keys that will be used for the other parts of the algorithm. Already described here in previous topics.

After tests and simulations, we obtain the following results compared to other methods in the Table 1:

Table 1 - Power Loss Compared to Other Algorithms

| Algorithm | Power Loss (MW) |
|---|---|
| Original Network (IEEE-14) | 13.436 |
| PSO Algorithm | 9.9159 |
| MPSO Algorithm | 8.5053 |
| ABC Algorithm | 6.4611 |
| FSS Algorithm | 7.8457 |
| HATSGA | 4.2796 |
| GSA Algorithm | 3.2764 |
| **ML+HATSGA** | **3.6796** |

IV - CONCLUSION AND FUTURE WORK

The proposed machine learning algorithm and the computational modeling in "R" language, proved to be satisfactory in support of decision making in reconfigurations of distribution networks in the context of Smart Grid discussed and illustrated during the work. The proposed algorithm uses artificial intelligence, machine learning and linear regression, as well as branch exchange in the search for a better solution for reconfiguration, producing initially relevant results to help improve the service in relation to the interruption of electricity supply and other anomalies. The new advantages achieved by the system are the ability to search for a better reconfiguration solution in a timely manner and the solution of reconfiguration of RDs in context of distributed generation and smart grids, fully adapted to the context and using graph theory and modeling in " R "with flexibility that can be expanded beyond the presented scenario. For future work we have, automatic preventive decision making in case of failure, example of a hospital transformer, simulate other IEEE topology test cases, analyze computational time spent searching for solutions.